# PIP-II TRANSFER LINES DESIGN*

A. Vivoli†, Fermilab, Batavia, IL 60510, USA

*Abstract*

The U.S. Particle Physics Project Prioritization Panel (P5) report encouraged the realization of Fermilab's Proton Improvement Plan II (PIP-II) to support future neutrino programs in the United States. PIP-II aims at enhancing the capabilities of the Fermilab existing accelerator complex while simultaneously providing a flexible platform for its future upgrades. The central part of PIP-II project is the construction of a new 800 MeV H⁻ Superconducting (SC) Linac together with upgrades of the Booster and Main Injector synchrotrons. New transfer lines will also be needed to deliver beam to the downstream accelerators and facilities. In this paper we present the recent development of the design of the transfer lines discussing the principles that guided their design, the constraints and requirements imposed by the existing accelerator complex and the following modifications implemented to comply with a better understanding of the limitations and further requirements that emerged during the development of the project.

## INTRODUCTION

PIP-II project envisions the construction of a new SC Linac, able to run in pulsed and CW mode, to increase the proton beam intensity available at Fermilab for its future experiments [1]. The project includes upgrades of the current synchrotrons chain (Booster, Recycler and Main Injector) and transfer lines to connect the new Linac to the rest of the accelerator complex. In particular, the main transfer line we present in this paper will transport the beam from the end of the SC Linac to the Booster. A second line is also presented to transport the beam to a possible upgrade of Mu2e experiment [2], to be built in the future.

A first design of the transfer lines was already presented [3], but a better understanding of the constraints and limitations raising with the development of the project made some modifications necessary. The main changes discussed here are the revision of the design of the transfer line to the Booster to resolve interferences with existing beamlines and infrastructures at Fermilab, design of the Dump line and revision of the switching system to support beam based energy stabilization and redirection of the beam to Mu2e future upgrade and consequent modifications of the transport line to Mu2e upgrade.

## TRANSPORT TO BOOSTER

A CAD drawing of the SC Linac, including ion source and warm front end, with the transfer lines on the picture of the Fermilab site is presented in Fig. 1.

The first part of the transfer line is reserved for the pos-sible future upgrade of the PIP-II Linac and is located in the straight extension of the Linac tunnel. The lattice consists of 4 doublet periods of the same length of the periods of the last section of the SC Linac, where cryomodules are installed between doublets, such that it is possible in the future to install additional cryomodules to increase the final energy of the SC Linac from 800 MeV to 1.2 GeV. The 4 doublets in this section can be used for matching the beam to the downstream line.

___________________________________________
* Work supported by Fermi Research Alliance, LLC under Contract No. DE-AC02-07CH11359 with the United States Department of Energy.
† vivoli@fnal.gov

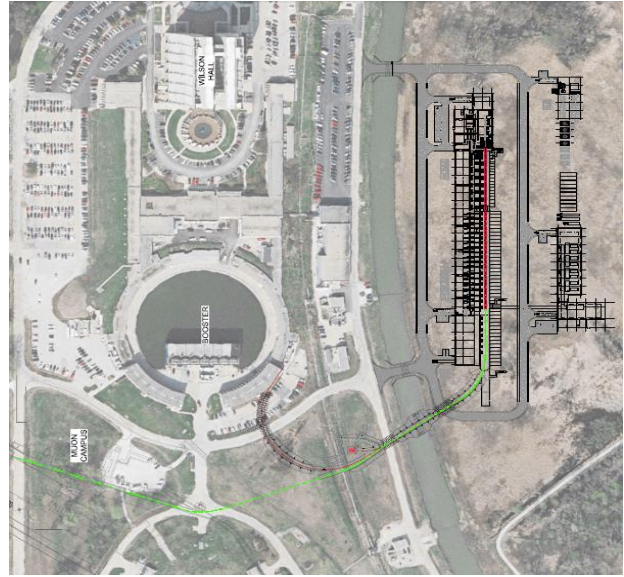

Figure 1: Drawing of SC Linac and transfer lines on the Fermilab site.

The injection into the Booster is vertical, with the final part of the transfer line being aligned with the injection straight of the Booster. At the beginning of this section the beam is brought at an elevation of 33.6 cm above the Booster plane by a magnetic dog-leg created by 2 dipole magnets and at the end of the section a vertical C-dipole bends the beam into the Booster injection magnets.

Between the energy upgrade and the injection sections the beam is transported by a FODO lattice. In the previous design the first 4 cells had dipoles between the quadrupoles forming the 1st arc, followed by 2 cells with no bends forming a straight transport and other 12 cells with bends forming the 2nd arc. The cell length was determined by the geometry of the accelerator complex and the strengths of the quadrupoles were chosen to have 90 deg. phase advance both horizontally and vertically. The dipole magnets were planned to be 2.45 m long and have a magnetic field of 2.36 kG. This was just below the value of 2.39 kG, fixed as a safe limit to avoid Lorentz stripping [4]. Note also that the choice of the phase advances per cell ensured both of the arcs to be achromatic.

With the development of the project, studies of possible interferences of the PIP-II design with existing accelerator infrastructure showed that the second arc would cross the Tevatron tunnel, where a transfer line bringing 120 GeV beams extracted from the Main Injector to the Fermilab Test Beam Facility is operating. This line is scheduled to be still in use at the time of PIP-II operations and its elevation is about 30 cm lower than the one of the transfer line, which is the same of the Booster and the SC Linac. After an accurate study of feasibility and cost of the options available it's been decided to keep the elevation of the SC Linac and transfer line the same of the Booster and create a vertical bump in proximity of the crossing, rising the line of about 1.3 m close to the ceiling of the Tevatron tunnel. This is done to leave the required 2.4 m of vertical clearance in the tunnel, needed for free passage of people and equipment.

From a lattice design point of view, it's been decided to create this vertical bump not increasing the number of dipoles but rolling them on their longitudinal axis, slightly increasing the magnetic field to keep the horizontal bending angle constant. Following this idea the first 5 dipoles of the second arc have been rolled to create the necessary elevation of the line and the following dipoles have been rolled in the opposite direction to bring the line back at the elevation of the SC Linac and Booster. This operation creates vertical dispersion and perturbs the horizontal one; to compensate these effects all the following dipoles have been rolled of some small angles to adjust the geometry of the line and make the second arc achromatic. To help satisfy the constraints also the vertical phase advance per cell has been changed to about 111 deg. and the magnetic field in the dipoles has been fixed at 2.41 kG. The final optical functions of the transfer line have been calculated with the code OptiM [5] and are presented in Fig. 2.

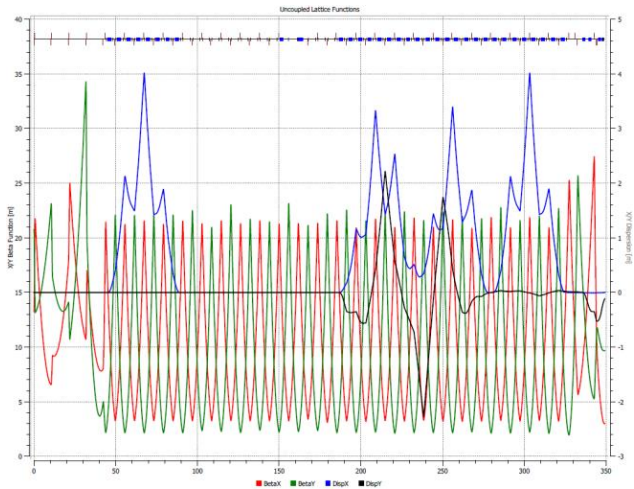

Figure 2: Optics of the transfer line from the SC Linac end to the stripping foil in the Booster.

Calculations carried out with OptiM also showed negligible x-y coupling.

## BEAM SWITCH DESIGN

The Dump line was originally planned at the end of the SC Linac in the extension of the Linac tunnel but it has been decided to move it to the straight section between the arcs where the switch to the Mu2e upgrade experiment was placed. For this reason the 2 switches have been replaced with only one system that can direct the beam to the Booster, to the Dump or to the Mu2e upgrade experiment (see Fig. 1). In the previous design the switches were realized with a vertical kicker used to send the beam off axis and a Lambertson magnet used to deflect it horizontally; a more accurate study showed that the design of the Lambertson magnet would be complicated so that now the switch is planned to be completely on the horizontal plane. A fast corrector is placed after the horizontally focusing quadrupole of the 6$^{th}$ cell in the straight transport section while the Lambertson magnet with 3 apertures is placed after the horizontally focusing quadrupole of the 7$^{th}$ cell. When the field is on in the fast corrector the beam is displaced offset from the axis of the vacuum chamber which will have an increasing cross section. Then the beam passes the next 2 large quadrupoles that deflect it into a side aperture of the Lambertson where the magnetic field sends it into the correct beamline. An invertion of the field polarity in the corrector sends the beam on a symmetric trajectory to a different beamline. When the field is off in the corrector the beam remains on the axis and enters the central aperture of the Lambertson where there is no field, so that it continues on a straight trajectory to the Booster beamline.

A simulation of the horizontal 10σ beam envelope in the switch with vacuum chamber and magnet apertures is presented in Fig. 3.

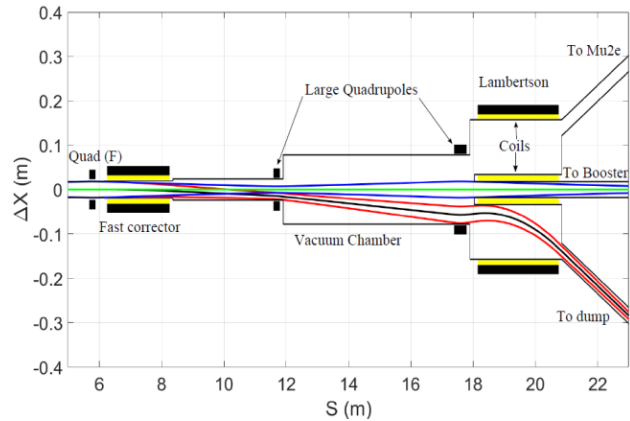

Figure 3: Simulation of the horizontal 10σ beam envelope going to the Booster (blue) and to the dump (red).

This design of the beam switch supports the beam based energy stabilization system that will be used in case the primary energy stabilization system, based on internal feedbacks and correction to the cavity voltages, will not guarantee the required level of stability.

## DUMP LINE

The decision to move the Dump line from the end of the SC Linac to the straight transport section between the arcs came from the necessity to keep clear the energy upgrade area from radiation, to leave a free passage of people and equipment in the SC Linac tunnel and to place the energy measurement system in the first arc. The dump will be able to dispose of up to 50 kW of the 800 MeV beam from the SC Linac and requires a 10 m clearance from other beamlines and equipment. To fulfil this constraint 5 dipole magnets of the same family used in the transport line to the Booster have been planned in the design of the Dump line, while focusing will be provided by 4 quadrupoles of the same family used in the FODO cells of the Booster line. A sweeping magnet in the long drift before the dump is planned to reduce the power density on the dump entrance. The optics functions of the Dump line have been calculated with OptiM and are showed in Fig. 4.

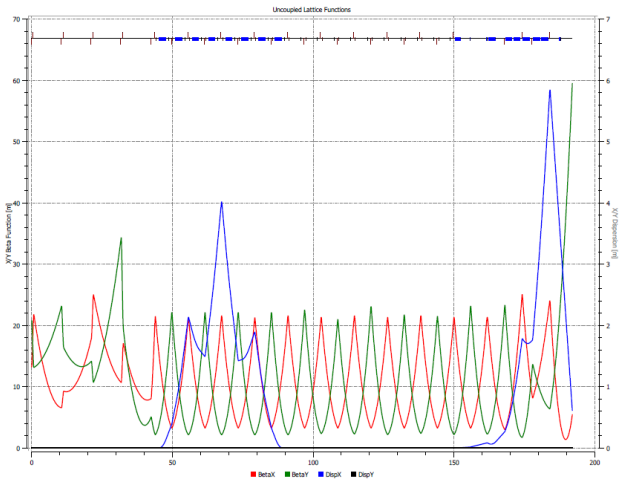

Figure 4: Optical functions of the line from the end of the SC Linac to the dump.

## TRANSPORT TO MU2E UPGRADE

The possibility to use the PIP-II beam for the Mu2e upgrade is one of the motivations for requiring the SC Linac to be able to run in CW mode. Studies show that the sensitivity of the Mu2e experiment could increase of a factor 10 if this upgrade is realized [2].

The transport line to the Mu2e upgrade will connect the beam switch system described above to the M4 line of the Mu2e accelerator chain. The idea is to join this external line with the same optics functions planned in the original design, reducing to a minimum the changes necessary for the beam transport to the upgrade.

The design of the transport line presents the problem of crossing the Tevatron ring like the line to the Booster but it also has to cross the MI-8 line, that brings the 8 GeV beam extracted from the Booster to the Recycler. To cross the Tevatron tunnel the same strategy used for the Booster line has been used, employing 4 dipole magnets of the same family used in the previous lines to bring the beam close to the tunnel ceiling. There are different options to be evaluated to decide the design downstream the crossing, but for the moment the line is not planned to go back to the elevation of the SC Linac/Booster, since line M4 has an elevation of 1.8 m above the Booster and the MI-8 line 3.3 m below it. So, it has been decided to keep the line elevation after the crossing constant and use a final arc to rise it to the junction point with line M4. The transfer line will pass MI-8 at an elevation of 4.5 m above it, which may be sufficient to keep their enclosures separated. Further studies are required to investigate feasibility and cost of this solution.

The final arc of the line is designed to use a different family of dipoles than the other lines since it would be difficult to fulfil the geometrical and optical constraints otherwise. The dipoles used have a magnetic field of 2.36 kG. For the same reason also this line lattice is made of FODO cells but with length and phase advances slightly different. As a consequence the gradient of the quadrupoles used in this section slightly differs from the values used for the transport to the Booster. Optics functions of this line have also been calculated with OptiM and are presented in Fig. 5.

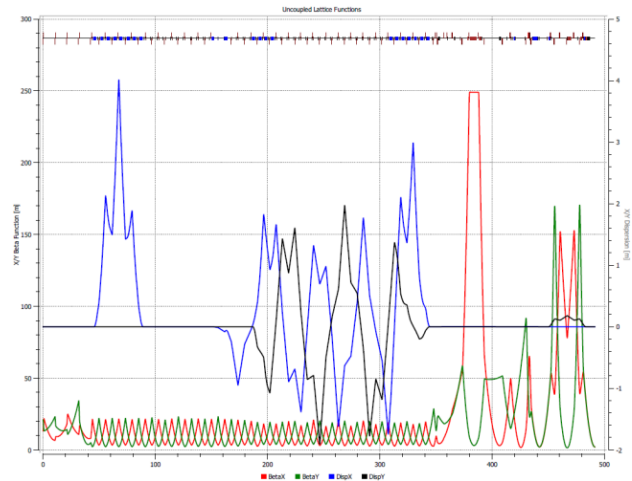

Figure 5: Optics of the line from the end of the SC Linac to the end of line M4.

## CONCLUSION

In this paper the new design of the PIP-II transfer lines have been presented. The work done is more or less finalized for the publication of the Conceptual Design Report of the project. More work and studies need to be done in view of the preparation of the Technical Design Report, including engineering study of the proposed solutions and accurate design of magnets and other elements required.